\documentclass[journal,twoside]{IEEEtran}

\usepackage{etoolbox}
\makeatletter
\@ifundefined{color@begingroup}%
  {\let\color@begingroup\relax
   \let\color@endgroup\relax}{}%
\def\fix@ieeecolor@hbox#1{%
  \hbox{\color@begingroup#1\color@endgroup}}
\patchcmd\@makecaption{\hbox}{\fix@ieeecolor@hbox}{}{\FAILED}
\patchcmd\@makecaption{\hbox}{\fix@ieeecolor@hbox}{}{\FAILED}

\usepackage{cite}
\usepackage{amsmath,amssymb,amsfonts}
\usepackage{lipsum}
\usepackage{graphicx}
\usepackage{textcomp}
\usepackage{siunitx}
\usepackage{float}
\usepackage{comment}
\usepackage{tabularx,booktabs}
\usepackage[ruled,linesnumbered]{algorithm2e}
\usepackage{algpseudocode}
\usepackage{ragged2e}
\usepackage[misc]{ifsym}
\usepackage{hyperref}

\newcolumntype{C}{>{\centering\arraybackslash}X} 
\def\BibTeX{{\rm B\kern-.05em{\sc i\kern-.025em b}\kern-.08em
    T\kern-.1667em\lower.7ex\hbox{E}\kern-.125emX}}

\newcommand{\name}{DIFR3CT}

\makeatletter
\newcommand*{\rom}[1]{\expandafter\@slowromancap\romannumeral #1@}
\makeatother

\begin{document}
\title{\name{}: Latent Diffusion for Probabilistic 3D CT Reconstruction from Few Planar X-Rays}

\author{Yiran Sun, Hana Baroudi, Tucker Netherton, Laurence Court, Osama Mawlawi, \\ Ashok Veeraraghavan, \IEEEmembership{Fellow, IEEE}, and Guha Balakrishnan
\thanks{This work has been submitted to the IEEE for possible publication. Copyright may be transferred without notice, after which this version may no longer be accessible.}
\thanks{Yiran Sun, Ashok Veeraraghavan and Guha Balakrishnan are with the Department of Electrical and Computer Engineering at Rice University, Houston, TX 77005 USA (e-mail: ys92@rice.edu; vashok@rice.edu; guha@rice.edu).}
\thanks{Hana Baroudi, Tucker Netherton and Laurence Court are with the Department of Radiation Physics at University of Texas MD Anderson Cancer Center, Houston, TX 77030 USA (e-mail: hbaroudi@mdanderson.org; tnetherton@mdanderson.org; lecourt@mdanderson.org).}
\thanks{Osama Mawlawi is with the Department of Imaging Physics at University of Texas MD Anderson Cancer Center, Houston, TX 77030 USA (e-mail: omawlawi@mdanderson.org).}
}

\maketitle

\begin{abstract}

Computed Tomography (CT) scans are the standard-of-care for the visualization and diagnosis of many clinical ailments, and are needed for the treatment planning of external beam radiotherapy. Unfortunately, the availability of CT scanners in low- and mid-resource settings is highly variable. Planar x-ray radiography units, in comparison, are far more prevalent, but can only provide limited 2D observations of the 3D anatomy. In this work we propose \name{}, a 3D latent diffusion model, that can generate a distribution of plausible CT volumes from one or few ($<$ 10)  planar x-ray observations. \name{} works by fusing 2D features from each x-ray into a joint 3D space, and performing diffusion conditioned on these fused features in a low-dimensional latent space. We conduct extensive experiments demonstrating that \name{} is better than recent sparse CT reconstruction baselines in terms of standard pixel-level (PSNR, SSIM) on both the public LIDC and in-house post-mastectomy CT datasets. We also show that \name{} supports uncertainty quantification via Monte Carlo sampling, which provides an opportunity to measure reconstruction reliability. Finally, we perform a preliminary pilot study evaluating \name{} for automated breast radiotherapy contouring and planning -- and demonstrate promising feasibility. Our code is available at \href{https://github.com/yransun/DIFR3CT}{https://github.com/yransun/DIFR3CT}.

\end{abstract}

\section{Introduction}
\label{sec:introduction}
\IEEEPARstart{C}{omputed} Tomography (CT) scans are the standard-of-care for the diagnosis and treatment of a range of patient disorders including fractures, heart disease, and cancer. CT scanners operate by rotating an x-ray source and detector panel around a patient lying on a bed, while acquiring several hundred 2D projection images. These projection images are in turn combined (``backprojected'') to reconstruct a 3D image volume of the body region. 

\begin{figure}[t!]
\includegraphics[width=\linewidth]{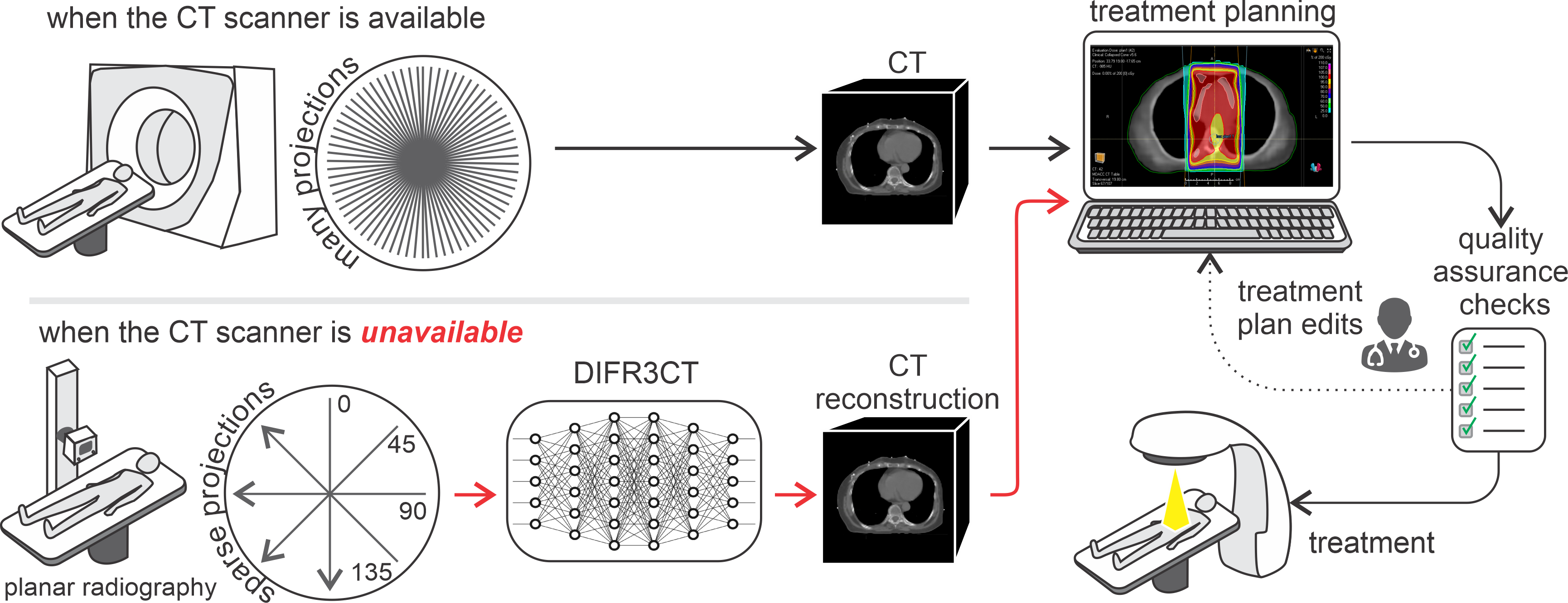}
\caption{\textbf{Extremely sparse-view CT reconstruction may be helpful in various low-resource settings for clinical applications like radiotherapy (RT) planning.} In the ideal RT pipeline (top), a CT scan is taken of a patient, and fed to a RT planning system. The resulting RT plan, which assigns doses to different anatomical regions, is examined and potentially corrected by a clinician, before being applied to the patient. When CT scanners are unavailable (bottom), we are interested in reconstructing the CTs to sufficient detail from extremely sparse planar x-ray images. We propose \name{} for such applications.}
\label{fig:RT-pipeline}
\end{figure}

Though they are valuable tools, CT scanners also have significant cost and infrastructure requirements, often making them infeasible for clinics without adequate resources or in impoverished areas. For example, a recent study indicates that there is less than 1 CT scanner per million inhabitants in low-to-middle-income countries (LMICs) compared to approximately 40 CT scanners per million inhabitants in high-income countries~\cite{hricak2021medical}. This lack of CT access directly impacts applications such as external beam radiotherapy (RT), which require 3D CT scans as input for state-of-the-art RT planning software tools (see Fig.~\ref{fig:RT-pipeline}).

%\begin{figure}[t!]
%\includegraphics[width=\linewidth, height=2.5in]{example-image-a}
%\caption{\textbf{\name{} takes as input one or more planar x-ray views and outputs a distribution of 3D CT scans consistent with these views.} \GB{Yiran, I suggest we make our teaser image more generic than RT planning. Basically, we can show some x-rays taken around a patient and fed into our model to output a distribution of potential scans.} \YS{Okay}}
%\label{fig:teaser}
%\end{figure}

%Increasing access to RT is a daunting challenge because LMICs have 80\% of the global cancer burden but only 5\% of global resources~\cite{atun2015expanding}. 
Compared to CT, planar radiography (2D projection) or ``x-ray'' imaging is a far cheaper and more widespread modality, particularly in LMICs~\cite{ngoya2016defining}. However, x-rays provide limited 2D anatomical information, and in few numbers are not sufficient to accurately reconstruct 3D anatomy using classical tomographic reconstruction methods. This precludes the use of x-rays in clinical applications which require accurate 3D anatomical information. Hence, a method that can accurately reconstruct CT scans under the \emph{extremely sparse} setting, i.e., $<10$ planar x-ray views, would have practical value in a number of applications such as RT planning in resource-constrained settings (see Fig.~\ref{fig:RT-pipeline}). %We denote this task \emph{extremely sparse-view CT reconstruction}, or \task{}.

Several recent studies demonstrate the promise of using deep neural networks to address extremely sparse CT reconstruction. Most use convolutional neural network (CNN) designs to fuse input x-rays together to predict a CT volume, supervised using pixel-wise reconstruction losses~\cite{kyung2023perspective,lin2023learning,shen2019patient,sun2023ct}, and adversarial losses~\cite{ying2019x2ct, cafaro2023x2vision}. While promising, these methods are predominantly \emph{deterministic}, i.e., provide only one reconstruction estimate, resulting in two shortcomings. First, because some high-frequency details are invariably lost in inverse imaging problems, a deterministic algorithm can ``smooth'' over unknown details resulting in low-resolution predictions. Second, these algorithms are incapable of providing \emph{uncertainty estimates} suggesting which regions of reconstructed CTs are susceptible to errors~\cite{barbano2022uncertainty}. More recent studies~\cite{liu2023dolce, chung2022improving, song2021solving} improve reconstruction details using diffusion models, the current state-of-the-art in \emph{probabilistic} deep generative modeling~\cite{dhariwal2021diffusion, ho2020denoising}. However, one significant limitation of these studies is that they generate 2D CT slices instead of full 3D CT volumes due to the high computational cost of training diffusion models on volumetric signals. \emph{In short, there is a need for a computationally efficient 3D diffusion model for extremely sparse CT reconstruction.}

To tackle the above issues, we propose \name{} (for Diffusion Reconstruction of 3D CT), a low-cost, probabilistic 3D CT reconstruction algorithm. The method takes one or more x-ray images as input, and outputs samples from the distribution of plausible 3D CT scans conditioned on a compact 3D representation of the x-rays. To lower computation costs, we build upon \emph{latent diffusion models}~\cite{rombach2022high}, by first learning a compact 3D ``latent'' space for the CT volumes, and then training a conditional diffusion model on top of this latent space. A key step in our approach is our design of the conditioning signal, which must combine information contained in the different input 2D x-rays into a coherent representation. To do so, we draw insights from the neural radiance fields (NeRF) algorithm family~\cite{mildenhall2021nerf, yu2021pixelnerf} by extracting 2D features from each x-ray and combining those features via ray tracing into a 3D feature volume based on the imaging acquisition geometry. \name{} can generate full 3D CT volumes with far reduced memory costs compared to vanilla 3D diffusion models \cite{ho2022video}, and enables computationally tractable uncertainty quantification in the form of posterior analysis through Monte Carlo sampling.

We evaluated \name{} on reconstructing CT scans from 1 to 8 input planar x-rays using two datasets: the public Lung Image Database Consortium (LIDC) CT dataset \cite{armato2011lung} and the in-house Thoracic post-mastectomy CT dataset (both datasets have roughly 1000 CT scans each). \name{} outperforms various sparse-view CT algorithms baselines in terms of voxel-level metrics (PSNR, SSIM~\cite{wang2004image}), and generates more convincing qualitative results. Second, we demonstrate that \name{} yields diverse realisations consistent with input x-rays, allowing for uncertainty quantification. Finally, using the Thoracic CT dataset, we conducted \emph{a first-of-its-kind case study} evaluating \name{} in the context of automated contouring and radiotherapy planning~\cite{Baroudi2023AutomatedCA} for 5 patients. We find that whole breast 2-field opposed radiotherapy plans using CTs reconstructed by \name{} meet dosimetric clinical goals for 3 out of the 5 plans. This demonstrates the potential feasibility of generating automate plans in settings where only planar imaging is available instead of volumetric CT. 

The contributions of this work are as follows:
\begin{enumerate}
  \item We propose \name{}, the first conditional latent diffusion model for high-quality extremely sparse CT reconstruction.
  \item We conduct experimental evaluations showing that \name{} outperforms state-of-the-art baselines in terms of PSNR and SSIM reconstruction accuracy metrics on LIDC and Thoracic CT datasets.
  \item We demonstrate that \name{} can provide reasonable uncertainty estimates for the reconstructed 3D CT scans. 
  \item We present the first application of an extremely sparse CT reconstruction algorithm towards a downstream clinical application: automated breast RT contouring and planning~\cite{Baroudi2023AutomatedCA}.
\end{enumerate}

\section{Related Work}
\subsection{Sparse-view Computed Tomography Reconstruction}
There are two broad types of sparse CT reconstruction tasks. The first, sparse-view CT reconstruction (SCTR), aims to reconstruct CTs from a few planar x-ray images taken at different orientations. The second, limited-angle CT reconstruction (LACTR), aims to reconstruct CTs from sinograms with limited input angles. Our work falls under SCTR, but particularly with extremely few ($<10$) views. Several recent studies use deep learning methods to address SCTR, which may be further divided into two categories: supervised models and generative models. 

Supervised SCTR models are typically implemented with convolutional neural network (CNN) and/or implicit neural representation (INR) network designs~\cite{lin2023learning, sun2023ct, kyung2023perspective, zha2022naf, ge2022x, kasten2020end}, and use mean squared or absolute error reconstruction losses. These algorithms predominantly suffer from over-smoothed results, due in part to their inability to handle ambiguities in the ill-posed reconstruction task. To alleviate this, one line of work uses patient-specific priors during the training stage~\cite{shen2019patient, shen2022nerp}, and others~\cite{ge2022x, sun2023ct} augment reconstruction losses with segmentation guidance.

The second group of SCTR models is based on deep generative modeling. Several methods build on generative adversarial networks (GANs)~\cite{cafaro2023x2vision,ying2019x2ct}. More recent studies use diffusion models, which tend to produce better outputs and distribution coverage than GANs. The basic idea of diffusion modeling is to gradually add Gaussian noise to a data distribution (known as the ``forward diffusion process''), and then learn to reverse it (``reverse diffusion process'') with deep neural networks (see Sec.~\ref{sec:background}). Diffusion models have been applied to both SCTR~\cite{chung2022improving,chung2023solving,lee2023improving} and LACTR~\cite{chung2022improving,chung2023solving,lee2023improving,liu2023dolce, guan2023generative} tasks. Unfortunately, diffusion models have high computational costs for volumetric data. Some studies address this by applying 2D diffusion models per 2D slice of a volume, and merging the results to reconstruct 3D CT volumes~\cite{chung2023solving,lee2023improving}, but this approach sacrifices some inter-slice consistency. In contrast, we design \name{} to predict entire 3D volumes at once using a compact latent space, using a latent diffusion model (LDM) framework~\cite{rombach2022high}. LDMs have been used in medical imaging for generation tasks of 3D CT and MRI~\cite{pinaya2022brain,khader2023denoising,zhu2023make}, but have yet to be widely used for SCTR.

\subsection{Radiotherapy Treatment Planning}
Radiotherapy (RT) treatment planning aims to prescribe an amount of radiation that can be safely administered to a targeted region of the body without injury to adjacent normal organs. %This treatment planning process takes days-to-weeks, after which the patient receives $>$20 separate appointments (for breast cancer). 
Before treatment planning, a patient receives a planning CT scan while lying in a position that will be exactly replicated during treatment. The physician determines how much radiation dose (measured in Gray [J/Kg]) to deliver to the target and uses a treatment planning system to volumetrically delineate the organs and tumor(s) and optimize dose delivery to irradiate the target and spare normal tissues. %After quality checks, digitally reconstructed radiographs (DRRs) are automatically created from the CT to depict the expected positioning of the patient during RT. 
Immediately prior to radiation delivery, the patient is placed on the table, and real-time radiographs acquired using linear accelerator (LINAC) onboard imaging are used to align the patient to the digitally reconstructed radiographs (DRRs) of the original CT for safe and accurate delivery of highly ionizing x-ray beams.

Crucially, if the CT scanner is unavailable (as can be the case in LMICs), the treatment can be either 1) delayed or 2) planned with hand calculations. For hand calculations, physical measurements and lookup tables are used to plan the treatment. Clinical evidence shows that 3D treatment planning decreases toxicity and increases local tumor control compared to hand calculations for many anatomical sites~\cite{huq2016report}. An accurate method to reconstruct CTs from a few radiographs can introduce a paradigm shift to eliminate delays and the use of hand calculations in planning (see Fig.~\ref{fig:RT-pipeline}). 
\section{Background on Diffusion Models}
\label{sec:background}

Denoising Diffusion Probabilistic Models (DDPMs) are powerful deep generative algorithms that achieve state-of-the-art performance for various generative tasks\cite{ho2020denoising, song2021scorebased}. Unconditional DDPMs approximate the true distribution of data samples using two processes: a fixed \emph{forward process} and a learning-based \textit{reverse process}. 

\textbf{Forward Process:} A fixed Markov chain that starts with a clean sample from the input data distribution $\mathbf{x}_0 \sim q(\mathbf{x}_0)$ and gradually adds Gaussian noise according to a variance schedule $\beta_{1:T}$, where $\beta_t \in (0, 1)$ for all $t \in [1, T]$:
\begin{equation}
    q(\mathbf{x}_t | \mathbf{x}_{t-1}) := \mathcal{N}(\mathbf{x}_t;\sqrt{1-\beta_t}\mathbf{x}_{t-1}, \beta_t\textbf{I})
    \label{eq:diff1}
\end{equation}
where $\mathbf{x}_T$ is an isotropic Gaussian distribution for large enough $T$ and a properly selected variance schedule. A nice property of this formulation is that we can also write $\mathbf{x}_t$ in closed form with respect to $\mathbf{x}_0$ directly, which allows for efficient training. %by optimizing random terms of loss function with stochastic gradient descent. 
Let $\alpha_t := 1-\beta_t$, $\bar{\alpha}_t := \prod^t_{s=1}\alpha_s$. Then we can sample $\mathbf{x}_t$ at any time step $t$ from:
\begin{equation}
    q(\mathbf{x}_t | \mathbf{x}_0) := \mathcal{N}(\mathbf{x}_t;\sqrt{\overline\alpha_t}\mathbf{x}_{0}, (1-\overline\alpha_t)\textbf{I})
    \label{eq:diff2}
\end{equation}
We can also rewrite Eq.~\ref{eq:diff2} as a linear combination of noise $\epsilon \sim \mathcal{N}(0, \textbf{I})$ and $\mathbf{x}_0$ as:
\begin{equation}
    \mathbf{x}_t = \sqrt{\overline{\alpha}_t}\mathbf{x}_0 + \sqrt{1-\overline{\alpha}_t}\epsilon
    \label{eq:diff3}
\end{equation}

\textbf{Reverse Process:} A joint Markov Chain distribution $p_{\theta}(\mathbf{x}_{0:T}) := p(\mathbf{x}_T)\prod_{t=1}^T p_{\theta}(\mathbf{x}_{t-1} | \mathbf{x}_t)$ with learned Gaussian transitions starting at $p(\mathbf{x}_T) = \mathcal{N}(\mathbf{x}_T; \textbf{0}, \textbf{I})$. We can learn the transition $p_\theta(\mathbf{x}_{t-1}|\mathbf{x}_{t})$ using a neural network $\mu_{\theta}(\cdot, \cdot)$:
\begin{equation}
    p_{\theta}(\mathbf{x}_{t-1} | \mathbf{x}_t) := \mathcal{N}(\mathbf{x}_{t-1}; \mu_{\theta}(\mathbf{x}_t, t), \textstyle\sum\nolimits_{\theta}(\mathbf{x}_t, t)),
\end{equation}
where $\theta$ represents the learnable parameters of the neural network. We can further reparameterize $\mu_{\theta}(\cdot, \cdot)$ by:
\begin{equation}
    \mu_{\theta}(\mathbf{x}_t,t) = \frac{1}{\sqrt{\alpha_t}}\left(\mathbf{x}_t - \frac{1-\alpha_t}{\sqrt{1-\bar{\alpha}_t}}\epsilon_{\theta}(\mathbf{x}_t,t)\right),
\end{equation}
where $\epsilon_{\theta}(\cdot, \cdot)$ predicts the noise added at each time step. The learning loss function of the $t$-th time step is then:
\begin{equation}
\mathcal{L}_t := \mathbb{E}_{t\sim [1,T],\mathbf{x}_0,\epsilon}\left[||\epsilon_t - \epsilon_{\theta}(\mathbf{x}_t,t)||^2\right]
\end{equation}
%During training, $n$ is randomly sampled from $\{0,1,...,N\}$, and the process is repeated until convergence. 
At inference time, given a sample of Gaussian noise \(\mathbf{x}_T \sim \mathcal{N}(0, \mathbf{I})\), we use $\epsilon_{\theta}(\cdot, \cdot)$ to progressively denoise $\mathbf{x}_T$ over \(T\) steps to generate a clean data point \(\mathbf{x}_0\). %Crucially, different input noise samples will result in different data samples, allowing one to simulate $p(\mathbf{x})$ via Monte Carlo sampling. 
%A U-Net architecture is commonly used to implement the denoising network for image data.

% Original Section for conditional diffusion models

\textbf{Conditional Diffusion Models (CDMs)} approximate $p(\mathbf{x} | \mathbf{c})$, where \(\mathbf{c} \in \mathbb{R}^{C}\) is some information describing the desired data sample, e.g. an attribute/caption if the data sample is an image. A denoising network \(\epsilon_{\theta} \colon \mathcal{X} \times \mathcal{C} \rightarrow \mathcal{X}\) 
%predicts a less noisy version of the input signal \(\mathbf{x} \in \mathbb{R}^{D}\),  
now also conditions the denoising steps on $\mathbf{c}$ using loss function:
% \GB{Yiran, check that this is equation is right, I modified this.}
%\begin{equation*}
%    \mathbf{x}_{t-1} = \dm(\mathbf{x}_t, t, \mathbf{c}) 
%    \text{\qquad for } t = T, \ldots, 1.
%\end{equation*}
\begin{equation}
\mathcal{L}^{CDM}_t := \mathbb{E}_{t\sim [1,T],\mathbf{x}_0,\epsilon}\left[||\epsilon_t - \epsilon_{\theta}(\mathbf{x}_t,t,\mathbf{c})||^2\right].
\end{equation}

The most popular strategy for training the conditional diffusion model is \emph{classifier-free guidance}~\cite{ho2021classifier}, which is a form of conditioning dropout: some percentage of the time, the conditioning information $\mathbf{c}$ is removed and replaced with a special input value representing the absence of conditioning information. The resulting model learns to capture both the conditional and unconditional distributions and their differences. Sampling is performed using a linear combination of the conditional and unconditional score estimates:
\begin{equation}
    \epsilon'_{\theta}(\mathbf{x}_t, \mathbf{c}, t) = (1+w)\epsilon_{\theta}(\mathbf{x}_t, \mathbf{c}, t) - w\epsilon_{\theta}(\mathbf{x}_t, t),
\end{equation}

\noindent with scalar $w$ controling their relative contributions.

\textbf{Latent Diffusion Models (LDMs)} perform the forward and reverse diffusion processes in a low-dimensional latent space using pretrained encoder $f_E(\cdot)$ and decoder $f_D(\cdot)$ functions, leading to the following conditional training objective:
\begin{equation}
\mathcal{L}^{LDM}_t := \mathbb{E}_{t\sim [1,T],f_E(\mathbf{x}_0),\epsilon}\left[||\epsilon_t - \epsilon_{\theta}(f_E(\mathbf{x}_t),t,\mathbf{c})||^2\right].
\end{equation}

During inference, the final denoised latent vector is passed to decoder $f_D(\cdot)$ to produce the image sample.

\section{Methods}
Let $X^i = \left\{X^i_1, \cdots, X^i_K \right\}$ denote $K$ input planar x-rays for patient $i$. Each x-ray $X^i_k \in \mathbb{R}^{1 \times h \times w}$ is a single-channel 2D image with resolution $h \times w$, generated with acquisition geometry (e.g., orientations, source-to-detector distances) $\theta_k$. We assume the same $K$ acquisition settings $\theta = \left\{\theta_1, \cdots, \theta_K\right\}$ for all patients. We denote the patient's ground truth CT scan by $Y^i \in \mathbb{R}^{1 \times d \times h \times w}$, where $d$ is the number of axial slices. 

\begin{figure*}[t!]
    \centering
    \includegraphics[width=\textwidth]{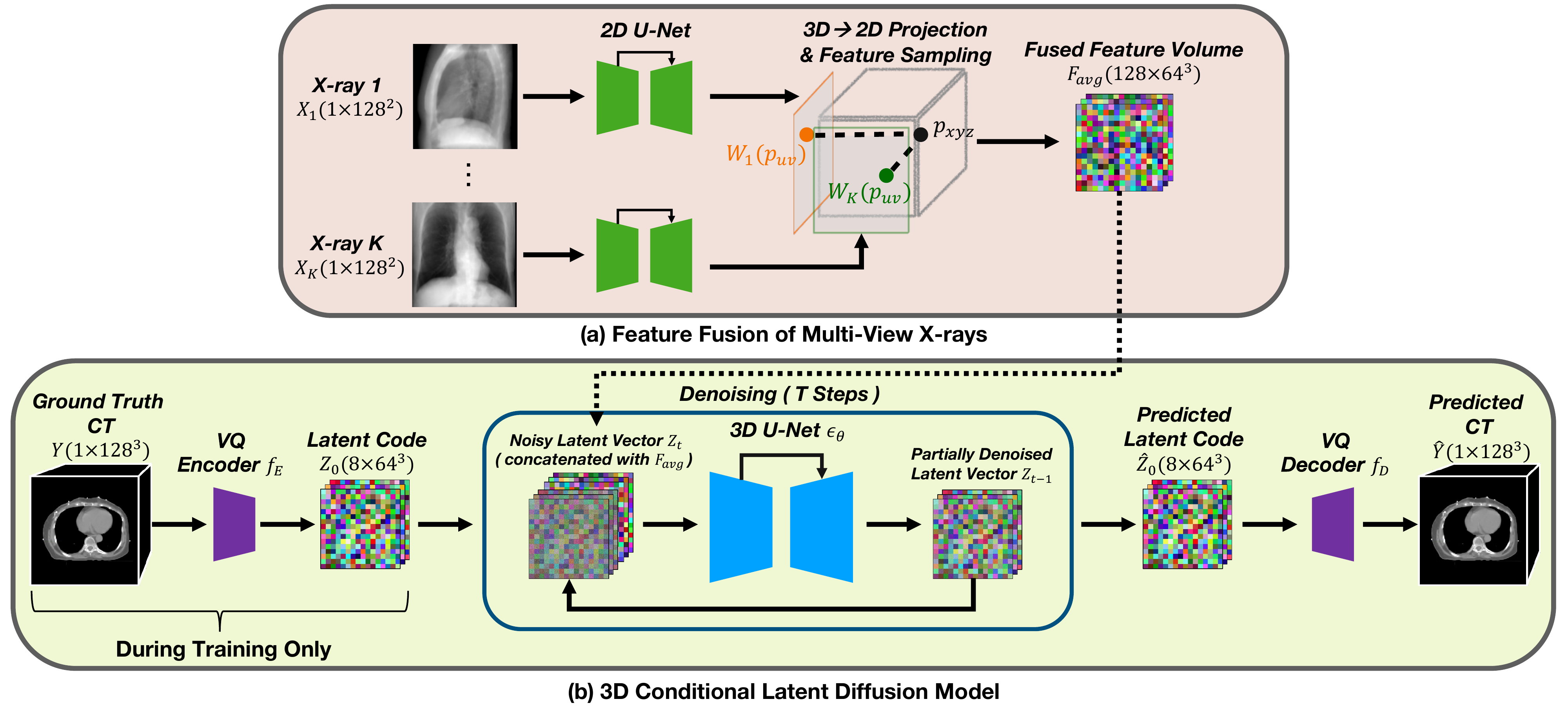}
    \caption{\textbf{Overview of \name{}.} \name{} consists of two parts. \textbf{a. Feature fusion of multi-view X-rays:} We extract a feature image $W_k$ from each input planar x-ray $X_k$ with a 2D U-Net. We then re-project $W_k$ back into 3D space using known x-ray imaging acquisition settings. We average all re-projected feature volumes into one feature volume $F_{avg}$. \textbf{b. 3D conditional latent diffusion model:} During training, each CT volume is encoded into a latent code $Z_0$ using a pretrained encoder~\cite{ge2022long}. We train a time-conditioned 3D denoising U-Net to take a random noisy latent code $Z_t$ and conditioning signal $F_{avg}$, and output a partially denoised code $Z_{t-1}$. After $T$ steps, the predicted code $\hat{Z}_0$ is reconstructed into a CT volume using a pretrained decoder.}
    %We also report 2D and 3D images of size in the form of \textit{c $\times$ hw} and \textit{c $\times$ dhw} separately.}
    % \GB{Need to adjust figure elements as we discussed: show x-rays in a circle at different angles, make feature fusion more geometric-ish, change bottom row so that diffusion is an iterative procedure, etc.}
    % \GB{Change camera picture on the right to something that looks more like an x-ray imaging system. Also, make sure that $Z_T$ looks like pure noise. Make sure that text inside the latent volumes is visible. You can make the CTs, latent noises, and stuff bigger if you reduce the whitespace between items in this figure. \YS{OK}}
    \label{fig:method1}
\end{figure*}

Our objective is to approximate and sample from the conditional distribution $p(Y^i|X^i; \theta)$. We develop \name{} to address this task, consisting of two components (see Fig.~\ref{fig:method1}): a feature fusion block (top) that constructs a joint 3D feature volume from the 2D x-rays, and a 3D conditional latent diffusion model (bottom) that operates over a learned 3D latent space of size significantly smaller than the CT volumes, resulting in computational savings with minimal sacrifice to reconstruction accuracy. We describe details of \name{} in the following sections.

\subsection{Feature Fusion of Multi-View X-rays}
\label{sec:feature_fusion}
We first combine the information from the input x-rays into one coherent feature space, which we will use as a conditional signal for our diffusion model (see Sec.~\ref{sec:ldm}). The main challenge to do so is that each x-ray is acquired with different acquisition geometry. Building on ray tracing ideas in Neural Attenuation Fields (NAF)~\cite{zha2022naf} and INRR3CT~\cite{sun2023ct}, we design \name{} to resample learned 2D features from each x-ray into one 3D volume using the known acquisition geometry (see Fig.~\ref{fig:method1}-top).

For each x-ray $X^i_{k}$, we first extract 2D features $W^i_{k} \in \mathbb{R}^{c \times h' \times w'}$ %via function $\mathbf{g_\psi}(\cdot)$ implemented 
with a 2D U-Net~\cite{ronneberger2015u}, where $c$ encodes the number of output features per pixel. Next, we construct one aggregate 3D feature volume from $W^i_{1}, \ldots, W^i_{K}$ with two steps: (1) resampling each $W^i_{k}$ into 3D space, and (2) averaging the resampled features across views. For the first step, let $\mathbf{p}_{xyz}\in\mathbb{R}^{1 \times 3}$ denote a 3D coordinate. We obtain the projected 2D coordinate $\mathbf{p}_{uv}\in\mathbb{R}^{1 \times 2}$ on $X^i_k$ by:
\begin{equation}
\mathbf{p}_{uv} = \mathcal{F}(\mathbf{p}_{xyz} \cdot R(\theta_k) + t),
\label{eq:geometry}
\end{equation}
where $\mathcal{F}(\cdot)$ is a fixed differentiable function that simulates the x-ray propagation process based on physical factors (Sec.~\ref{sec:tigre} details several of these factors for a common x-ray simulator), the most important being the projection type (parallel or cone beam)~\cite{long20103d, shen2022geometry}. In parallel radiation, each 3D point projects onto a 2D plane along parallel rays, and in cone-beam radiation, each 3D point projects onto a 2D plane based on rays emanating from a 3D source point. $R(\theta_k) \in \mathbb{R}^{3 \times 3}$ is the rotation matrix of angle $\theta_k$, and $t \in \mathbb{R}^{1 \times 3}$ is a translation matrix. Using Eq.~\eqref{eq:geometry}, we project all 3D points to their corresponding 2D locations in $W^i_{k}$, and use bilinear interpolation to extract their associated feature vectors, to produce feature volume $F^i_{k} \in \mathbb{R}^{{c \times d' \times h' \times w'}}$. Finally, in step 2, we aggregate the $K$ feature volumes into one volume via element-wise average pooling: $F^i_{avg} = \frac{1}{K}\sum^K_{n=1}(F^i_{k})$. We use $F^i_{avg}$ as the conditioning signal to the diffusion model.

\subsection{3D Conditional Latent Diffusion Model (LDM)}
\label{sec:ldm}
%Our LDM operates over a compact, learned 3D latent space, and a 3D diffusion architecture. We describe these in detail in this section.

\subsubsection{Learning the Latent Space}
A good latent space should capture important semantic factors of the CT data distribution, while attenuating imperceptible, high-frequency spatial details. We choose to construct this space using VQGAN~\cite{ge2022long}, which has demonstrated successful image encoding ability for applications such as text-to-image generation~\cite{rombach2022high}. 

We train one 3D VQGAN model per training CT distribution, consisting of an encoder $f_E(\cdot)$ and a decoder $f_D(\cdot)$. The encoder converts a CT volume $Y^i \in \mathbb{R}^{1 \times d \times h \times w}$ into a latent code $Z_0^i \in \mathbb{R}^{1 \times d' \times h' \times w'}$, where $dhw > d'h'w'$. In our experiments, $dhw = 128^3$, and $d'h'w'= 64^3$. The decoder $f_D(\cdot)$ reconstructs the CT volume $\hat{Y}^i$ from $Z_0^i$. 

VQGAN uses several training loss functions, which we tailor for our application. The first, $\mathcal{L}_{VQVAE}$, is identical to the one used in VQVAE~\cite{van2017neural}, consisting of reconstruction and KL-divergence regularization terms. We also add two adversarial losses ($\mathcal{L}_{D3}$, $\mathcal{L}_{D2}$) and a perceptual loss ($\mathcal{L}_{P}$) to promote realistic reconstruction details. The two adversarial losses have the form: 
\begin{equation}
    \mathcal{L}_{D3}(Y,\hat{Y}) = h(1-D3(Y)) + h(1+D3(\hat{Y})),
\end{equation}
\begin{equation}
    \mathcal{L}_{D2}(Y,\hat{Y}, s) = h(1-D2(Y[s]))+h(1+D2(\hat{Y}[s])),
\end{equation}
% \begin{comment}
% \begin{align}
% \mathcal{L}_{D3}(Y,\hat{Y}) &= \max(0, 1-D_3(Y)) + \max(0, 1+D_3(\hat{Y}))
% \end{align}
% \begin{multline}
%     L_{D2}(Y,\hat{Y}) = \sum_{i \in S} \left(\max(0, 1-D_2(Y[i, :, :])) \\
%                            + \max(0, 1+D_2(\hat{Y}[i, :, :])) \right)
% \end{multline}
% \end{comment}
where $h(x) = \max(0, x)$ is the hinge function~\cite{lim2017geometric}, and $D2$ and $D3$ are ``discriminator'' networks. $D3$ predicts whether a 3D volume belongs to the true distribution of CT volumes, and $D2$ predicts whether a 2D image belongs to the true distribution of axial CT slices, where $Y[s]$ indexes an axial slice of CT volume $Y$ at index $s$. We apply a perceptual loss $\mathcal{L}_{P}$~\cite{wang2018high} evaluating reconstructed 2D axial slices with respect to ground truth slices in the activation space of the VGG16~\cite{simonyan2014very} network pretrained on ImageNet~\cite{deng2009imagenet}.
%Let $f^{j}_{VGG}(\cdot, \cdot)$ represent the $j$th layer of a VGG16~\cite{simonyan2014very} network pretrained on ImageNet~\cite{deng2009imagenet}. Then:
%\begin{equation}
%\mathcal{L}_{P}(Y, \hat{Y}, s) = \sum_{j}p_j\|f^{j}_{VGG}(\hat{Y}[s])-f^{j}_{VGG}(Y[s])\|_1
%\end{equation}
%where $p_j$ is a fixed scaling factor per-level. 

The overall VQGAN training objectives for the discriminators and generator are:
\begin{align}
\mathcal{L}_{D} &= \arg\min[\lambda_1 \cdot (\mathcal{L}_{D3} + \mathcal{L}_{D2})] \\
\mathcal{L}_{G} &= \arg\min[\lambda_2 \cdot \mathcal{L}_{VQVAE} + \lambda_3 \cdot \mathcal{L}_P]
\end{align}
where $\lambda_1$, $\lambda_2$, $\lambda_3$ control the importance of each loss term.

\subsubsection{Conditional Diffusion} 
In line with most existing diffusion studies, we use a time-conditioned U-Net architecture~\cite{ho2020denoising, rombach2022high} to perform denoising at each time step of the inverse diffusion process (see Fig.~\ref{fig:method1}-bottom). We incorporate the conditioning signal $F_{avg}$ into the denoising process by simply concatenating it to the noisy target latent codes in the channels dimension as input to the U-Net. During each iteration of the training process, we randomly select a CT volume with its associated x-rays, and train the U-Net using the loss function:
\begin{equation}
    \mathcal{L}_{LDM} := \mathbb{E}_{t, Z_0, \epsilon, F_{avg}}[\|\epsilon_t - \epsilon_{\theta}(Z_t, t, F_{avg})\|^2].
\end{equation}

We train the U-Net using classifier-free guidance (see Sec. \ref{sec:background}-CDMs). Specifically, we jointly train a single LDM on both conditional and unconditional objectives by randomly dropping $F_{avg}$ (i.e., setting it to 0). In our experiments, we use $T=1000$ diffusion timesteps during training. However, during inference, we use \textit{DPM-SOLVER++}~\cite{lu2022dpm}, a sampler which can achieve high-resolution synthesis in only $T=10$ steps without needing to retrain or fine-tune the model, resulting in a significant inference speedup. 

% We then perform sampling using the following linear combination of the conditional and unconditional score estimates, where $w$ denotes the guidance strength:
% \begin{equation}
%     \epsilon'_{\theta}(Z_t, F_{avg}, t) = (1+w)\epsilon_{\theta}(Z_t, F_{avg}, t) - w\epsilon_{\theta}(Z_t, t)
% \end{equation}

%\subsection{Model Architecture and Inference Details}
%\GB{This paragraph seems unnecessary, doesn't add much.}The network architecture of \name{} is similar to the original configurations of the 3D convolutional VQGAN and 3D diffusion model.  The convolutional VQGAN involves training a codebook with visually rich context and modeling their composition with an autoregressive transformer architecture. It is trained alongside 2D and 3D patch discriminators, adapted from PatchGANs~\cite{isola2017image}, to distinguish real from fake 2D axial slices and 3D CT volumes, thereby retaining high perceptual quality.

\subsection{Uncertainty Estimation}
\label{sec:uq}
% \GB{Update this section to mention bias/variance}. 
Uncertainty quantification is key to building trustworthy AI systems for clinical applications. A key benefit of using diffusion models for reconstruction tasks is that we can naturally perform statistical analysis via Monte Carlo (MC) sampling. 

Uncertainty in inverse problems can be broadly divided into \textit{aleatoric} and \textit{epistemic} types~\cite{kendall2017uncertainties}. Aleatoric uncertainty pertains to variabilities caused by fundamental random factors of an inverse problem, while epistemic uncertainty pertains to variabilities caused by the inference model’s lack of knowledge
or understanding, which can be reduced with more diverse training data~\cite{Chan2024HyperDiffusionEE, Hllermeier2019AleatoricAE}. The aleatoric uncertainty of the reconstruction task can be captured by the \emph{variance} of image features over multiple predicted CT samples from the distribution $p(\cdot |X^i; \theta)$, learned by \name{}. Sample variance converges to the true \textit{aleatoric} uncertainty as $N \rightarrow \infty$~\cite{Chan2024HyperDiffusionEE}.

% \GB{Yiran, Fill in the equation}

During inference for patient $i$, we perform MC sampling by generating \textit{N} random CT samples $\mathcal{Y}^i = \{\hat{Y}^i_0, \hat{Y}^i_1,..., \hat{Y}^i_N\}$ from \textit{N} random noise codes $\{Z_0, Z_1,..., Z_N\}$. Using these samples, we decompose the per-voxel error of \name{} on this patient into bias and variance components: $Error(\mathcal{Y}^i)^2 = Bias(\mathcal{Y}^i)^2 + Var(\mathcal{Y}^i)$~\cite{tibshirani1996bias}, where Bias captures the average error per voxel: $Bias = \frac{1}{N} \sum_{n=1}^{N} (\hat{Y}^i_n - Y^i_n)$, and $Var$ captures the per-voxel variance. Bias computation requires a ground truth CT, which will not be available at inference time in the clinical settings targeted by \name{}. However, bias is a useful metric to analyze during \emph{model development} to assess and ensure the responsible use of the model~\cite{mehrabi2021survey}.

%Therefore, while bias is not possible to compute in a clinical setting without ground truth, it is a useful metric for model development.
%we can produce voxel-wise mean prediction, variance, squared bias maps (remember that the ground truth CT is not known during real clinical practice). Bias and variance are crucial in analyzing uncertainty because they reveal which parts of an image are more likely to introduce realistic artifacts compared to the ground truth. 

\section{Experiments}
We qualitatively and quantitatively evaluated \name{}'s performance conditioned on different numbers of input x-ray views. For each number of views, we trained a separate instance of \name{}. We compare \name{} against four learning-based CT reconstruction baselines (see Sec.~\ref{sec:recon_results}). We evaluated reconstruction accuracies using classical metrics (PSNR and SSIM), and radiotherapy dose volume histogram metrics (e.g. V90\%, V20Gy) (see Sec.~\ref{sec:metrics}). For the latter, we used a previously developed automated contouring tool to segment all tissues (e.g. breast, heart, lung)~\cite{baroudi2023automated}, and an automated radiotherapy planning tool~\cite{aggarwal2023radiation} to create dose distributions. Our code for reproducing results in this section is available at \href{https://github.com/yransun/DIFR3CT}{https://github.com/yransun/DIFR3CT}.

\begin{figure}[t!]
    \centering
    \includegraphics[width=\linewidth]{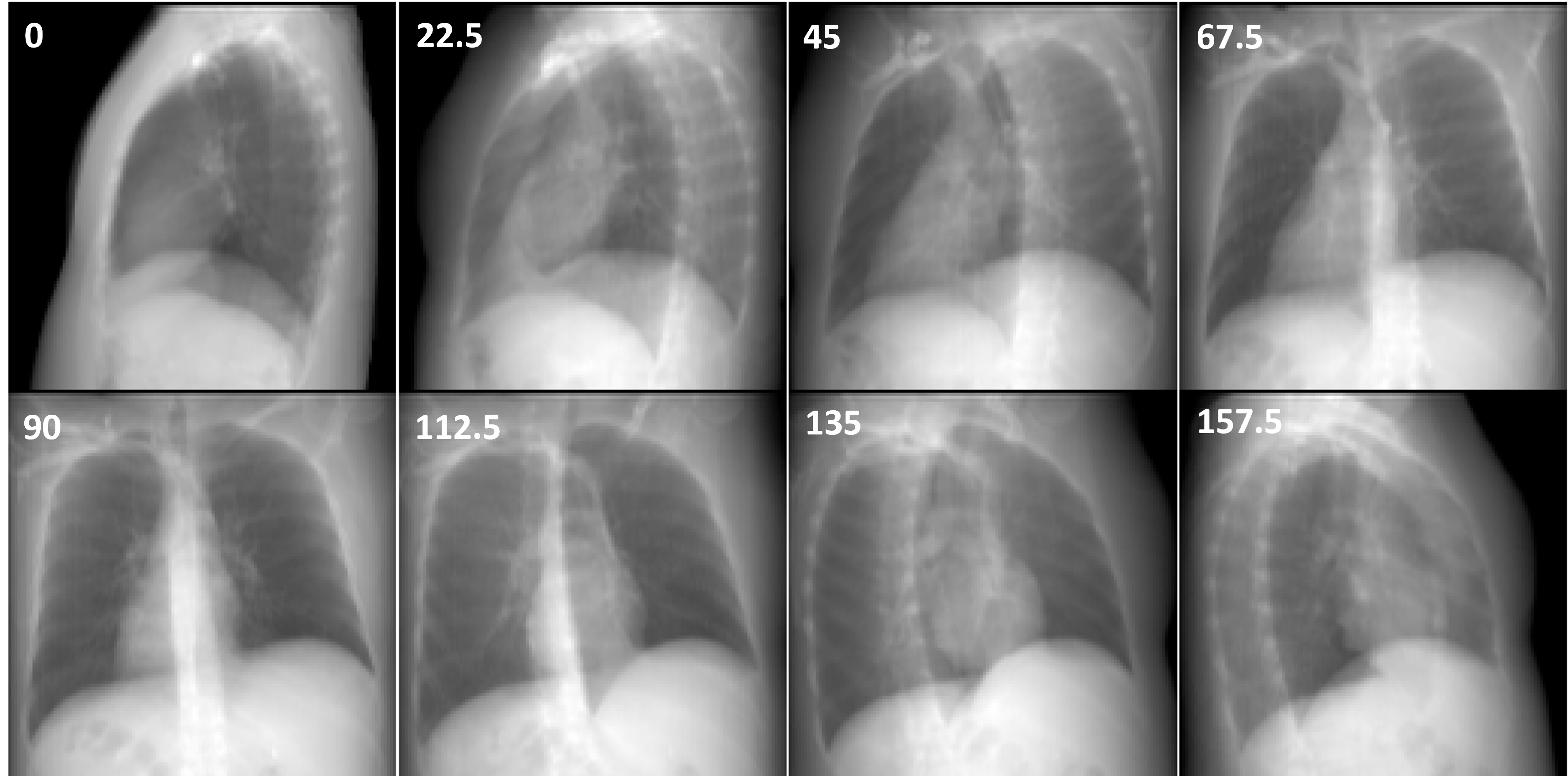}
    \caption{\textbf{Example x-rays generated by the TIGRE~\cite{biguri2016tigre} DRR generator for one LIDC CT volume.} We generated these x-rays at eight angles (printed on the top-left corner of each x-ray in degrees) around the CT volume.}
    \label{fig:TIGRE}
\end{figure}

\begin{table*}[t!]
\caption{\textbf{Quantitative evaluation of all models on the LIDC CT Dataset, using PSNR and SSIM metrics.} A cell is marked with an 'X' when the corresponding number of views is not computationally feasible or possible to run with that particular model.}
\label{lidc}
\resizebox{\textwidth}{!}{%
\begin{tabularx}{\textwidth}{@{} l *{5}{C} c @{}}
\toprule
\textbf{Method} & \textbf{8-view} & \textbf{4-view} & \textbf{2-view} & \textbf{1-view (frontal)} & \textbf{1-view (lateral)}\\ 
Metric & PSNR$\uparrow$/SSIM$\uparrow$ & PSNR$\uparrow$/SSIM$\uparrow$ & PSNR$\uparrow$/SSIM$\uparrow$ & PSNR$\uparrow$/SSIM$\uparrow$ &  PSNR$\uparrow$/SSIM$\uparrow$ \\
Scan Angle & $0^\circ, 22.5^\circ, \ldots, 157.5^\circ$ & $0^\circ, 45^\circ, 90^\circ, 135^\circ$ &  $0^\circ, 90^\circ$ & $90^\circ$ & $0^\circ$ & \\

\midrule
\name{} (Ours) & \textbf{30.36/0.782} & \textbf{29.56/0.745} & \textbf{28.08}/0.699 & 22.27/0.485 & 24.94/0.593 \\ 
\midrule
INRR3CT \cite{sun2023ct}  & \scalebox{1.2}{$\times$} & 28.20/0.703 & 28.04/\textbf{0.702} & \textbf{23.22/0.545} & \textbf{26.07/0.663} \\
X2CT-GAN \cite{ying2019x2ct} & \scalebox{1.2}{$\times$} & \scalebox{1.2}{$\times$} & 26.59/0.639 & 22.11/0.476 & 23.99/0.548\\ 
3D Diffusion \cite{ho2022video} & \scalebox{1.2}{$\times$}  & \scalebox{1.2}{$\times$} & 24.24/0.443 & 18.81/0.264 & 23.69/0.522\\ 
NAF \cite{zha2022naf} (parallel beam) & 25.78/0.583 & 21.38/0.380 & 20.50/0.316 & 18.64/0.262 & 19.79/0.271\\  
\bottomrule
\end{tabularx}}
\label{tbl:lidc-results}
\end{table*}

\begin{table*}[t!]
\caption{\textbf{Quantitative evaluation of all models on the Thoracic CT Dataset, using PSNR and SSIM metrics.} A cell is marked with an 'X' when the corresponding number of views is not computationally feasible or possible to run with that particular model.}
\label{breast}
\resizebox{\textwidth}{!}{%
\begin{tabularx}{\textwidth}{@{} l *{5}{C} c @{}}
\toprule
\textbf{Method} & \textbf{8-view} & \textbf{4-view} & \textbf{2-view} & \textbf{1-view (frontal)} & \textbf{1-view (lateral)}\\ 
Metric & PSNR$\uparrow$/SSIM$\uparrow$ & PSNR$\uparrow$/SSIM$\uparrow$ &  PSNR$\uparrow$/SSIM$\uparrow$ & PSNR$\uparrow$/SSIM$\uparrow$ &  PSNR$\uparrow$/SSIM$\uparrow$ \\ 
Scan Angle & $0^\circ, 22.5^\circ, \ldots, 157.5^\circ$ & $0^\circ, 45^\circ, 90^\circ, 135^\circ$ &  $0^\circ, 90^\circ$ & $90^\circ$ & $0^\circ$ & \\

\midrule
\name{} (Ours) & \textbf{29.84/0.794} & \textbf{28.67/0.772} & \textbf{26.98/0.730} & 18.27/0.399 & 20.44/0.526\\ 
\midrule
INRR3CT \cite{sun2023ct} & \scalebox{1.2}{$\times$} & 26.20/0.684 & 25.62/0.675 & \textbf{19.58/0.443} & \textbf{22.90/0.623} \\
X2CT-GAN \cite{ying2019x2ct} & \scalebox{1.2}{$\times$} & \scalebox{1.2}{$\times$} & 24.66/0.643 & 17.80/0.371 & 19.95/0.477\\ 
3D Diffusion \cite{ho2022video} & \scalebox{1.2}{$\times$} & \scalebox{1.2}{$\times$} & 21.61/0.529 & 16.69/0.275 & 18.22/0.390\\ 
NAF \cite{zha2022naf} (parallel beam) & 26.10/0.645 & 21.68/0.571 & 19.79/0.532 & 16.49/0.338 & 17.83/0.465\\ 
\bottomrule
\end{tabularx}}
\label{tbl:thoracic-results}
\end{table*}

\subsection{Datasets and Preprocessing}
We used the public Lung Image Database Consortium (\emph{LIDC}) CT dataset~\cite{armato2011lung} and \emph{Thoracic}, an in-house chest wall CT dataset with patients who received mastectomy (gathered under an IRB-approved protocol). The LIDC CT dataset includes 1018 patients, which we randomly split into 868/50/100 train/validation/test groups, while the Thoracic CT dataset includes 997 patients, which we randomly split into 850/47/100 train/validation/test groups. We clipped all voxel values of lung CTs to $[0, 2500]$ Hounsfield Units (HU) and thoracic CTs to $[-1000, 1000]$ HU. We normalized all CT voxel values to the range $[0, 1]$ before training all models. We resampled each scan to $1$ mm$^3$ resolution, cropped the result to a cube and resized to $128^3$ voxels. 

\subsection{Digitally Reconstructed Radiograph (DRR) Generator} 
\label{sec:tigre}
We generated eight planar x-ray views per CT at the following angles: $0^{\circ}$ (Lateral), $22.5^{\circ}$, $45^{\circ}$, $67.5^{\circ}$, $90^{\circ}$ (Frontal), $112.5^{\circ}$, $135^{\circ}$, and $157.5^{\circ}$ using the DRR generator TIGRE~\cite{biguri2016tigre} (Fig.~\ref{fig:TIGRE}). TIGRE permits the setting of several image acquisition parameters, including distance between source and volume center (DSO), distance between source and detector plane (DSD), physical size of patient voxels (dVoxel), number of detector pixels (dDetector), and projection type (e.g., parallel or cone beam). We used the following settings for all experiments: DSO=1000\unit{mm}, DSD=1500\unit{mm}, dVoxel=1\unit{mm}, dDetector=1\unit{mm}, and parallel beam projections.

\subsection{Training and Implementation Details}
\textbf{\name{}}. We trained and evaluated our models using PyTorch~\cite{paszke2019pytorch} on NVIDIA A100 GPUs each with 40 GB of memory. We chose a compression factor of $2^3$ for our 3D VQGAN models (i.e. image of size $128^3$ have a latent size of $64^3$), with a codebook size and dimensionality of 4096 and 8, using the Adam optimizer with a fixed learning rate of $3 \times 10^{-5}$. 

We used $T=1000$ training timesteps for our LDMs and a linear noise schedule~\cite{chen2023importance}. We used the Adam optimizer with a fixed learning rate of $1 \times 10^{-4}$. We used a batch size of 1 and trained each model for 500 epochs, which took approximately 4 days using our setup. We simply selected the final model checkpoints for all experiments without any optimized model selection strategies.

\textbf{Baselines}. We compared \name{} to four baselines: 3D Diffusion~\cite{ho2022video}, X2CT-GAN~\cite{ying2019x2ct}, NAF~\cite{zha2022naf}, and INRR3CT~\cite{sun2023ct}. 3D Diffusion operates directly on 3D data without using a latent space. X2CT-GAN is a generative adversarial network (GAN) framework to reconstruct CT images from one or two orthogonal x-rays. NAF is a fast self-supervised method for sparse-view CT reconstruction based on neural rendering and implicit neural representations (INRs), and works well given around 50 input views. INRR3CT is a neural network meant for highly sparse number of views like \name{}, but is based on more conventional CNN and INR network architectures. Due to GPU memory constraints, we only used four 8-dimension self-attention ``heads'' per transformer block for 3D Diffusion models (as opposed to eight 32-dimension self-attention ``heads'' for \name{}), and only up to four input views for INRR3CT models. We used the public training configurations for all other baselines. %\GB{as opposed to how many for \name{}?}

% LIDC

\begin{figure*}[t!]
    \centering
    \includegraphics[width=\textwidth]{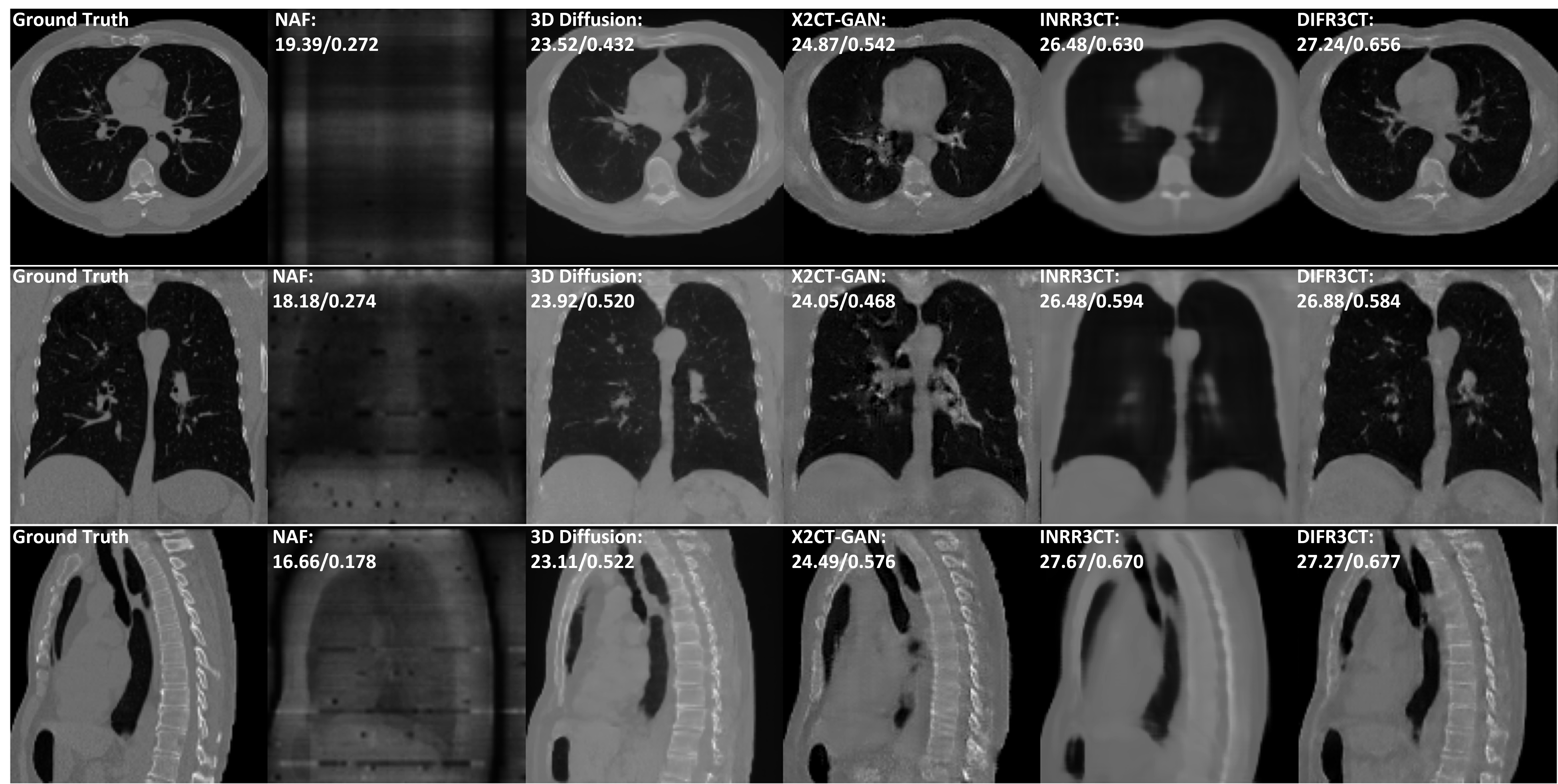}
    \caption{\textbf{Comparison of \name{} with baselines on the LIDC Dataset, given biplanar x-ray inputs.} Each row corresponds to a different center planar view of the CT volume (axial, coronal, sagittal). The second to fifth column correspond to four baselines (marked in text on each image), and the final column shows the reconstructed 3D CT images by using the proposed \name{} method. We also report PSNR/SSIM values on each slice. DIFR3CT generates the most realistic reconstructed details of all methods.}
    \label{fig:comlung}
\end{figure*}

\begin{figure*}[t!]
    \centering
    \includegraphics[width=\textwidth]{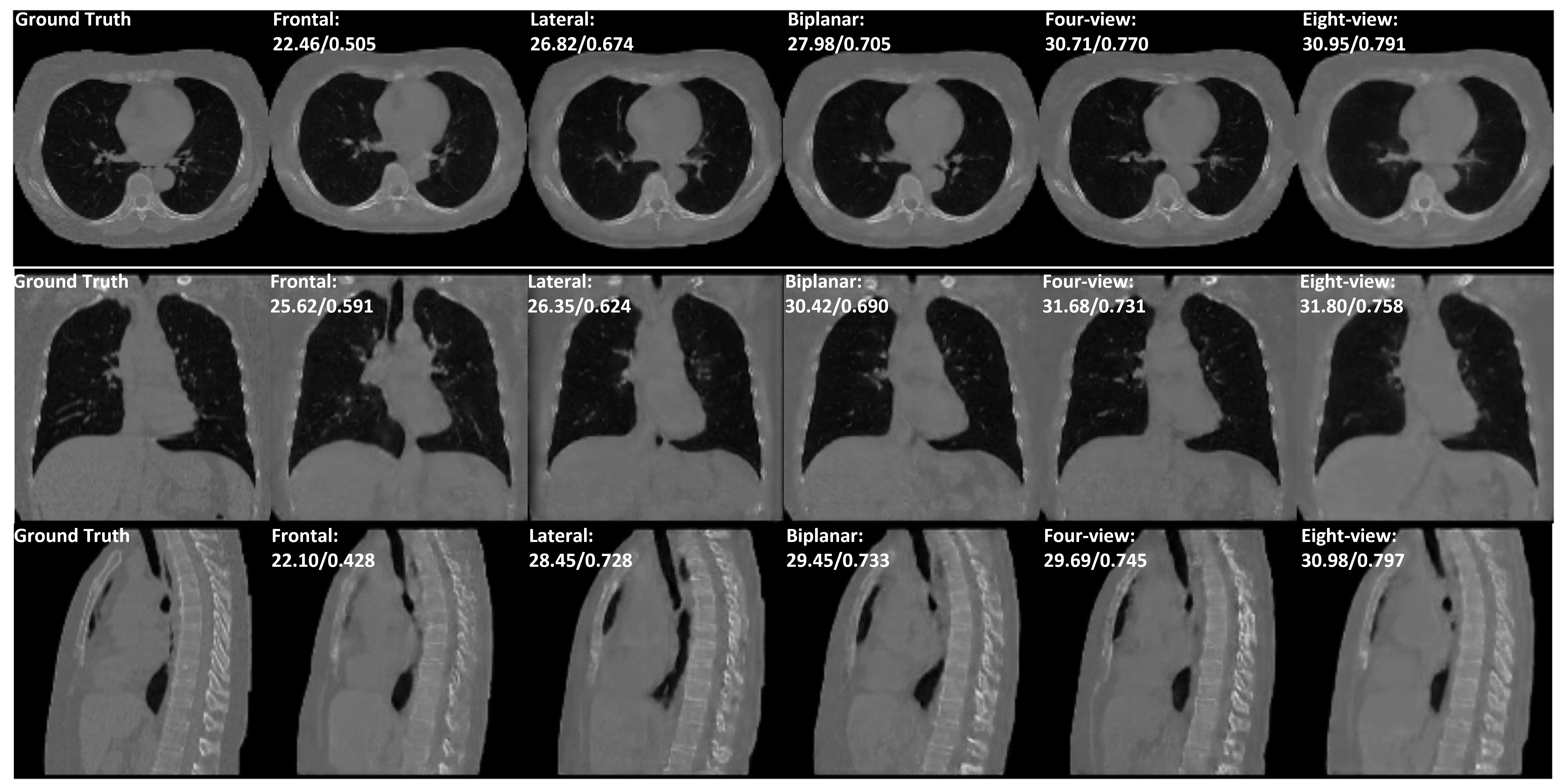}
    \caption{\textbf{Example 3D LIDC CT reconstruction results on one patient, with varying numbers of input views.} Each row corresponds to a different center planar view of the CT volume (axial, coronal, sagittal), and the second to sixth columns correspond to a different number of views (marked in text on each image). We also report PSNR/SSIM values on each slice. As the number of input viewing angles increases, the reconstruction details improve, especially near anatomical boundaries.}
    \label{fig:lung}
\end{figure*}

% Thoracic

\begin{figure*}[t!]
    \centering
    \includegraphics[width=\textwidth]{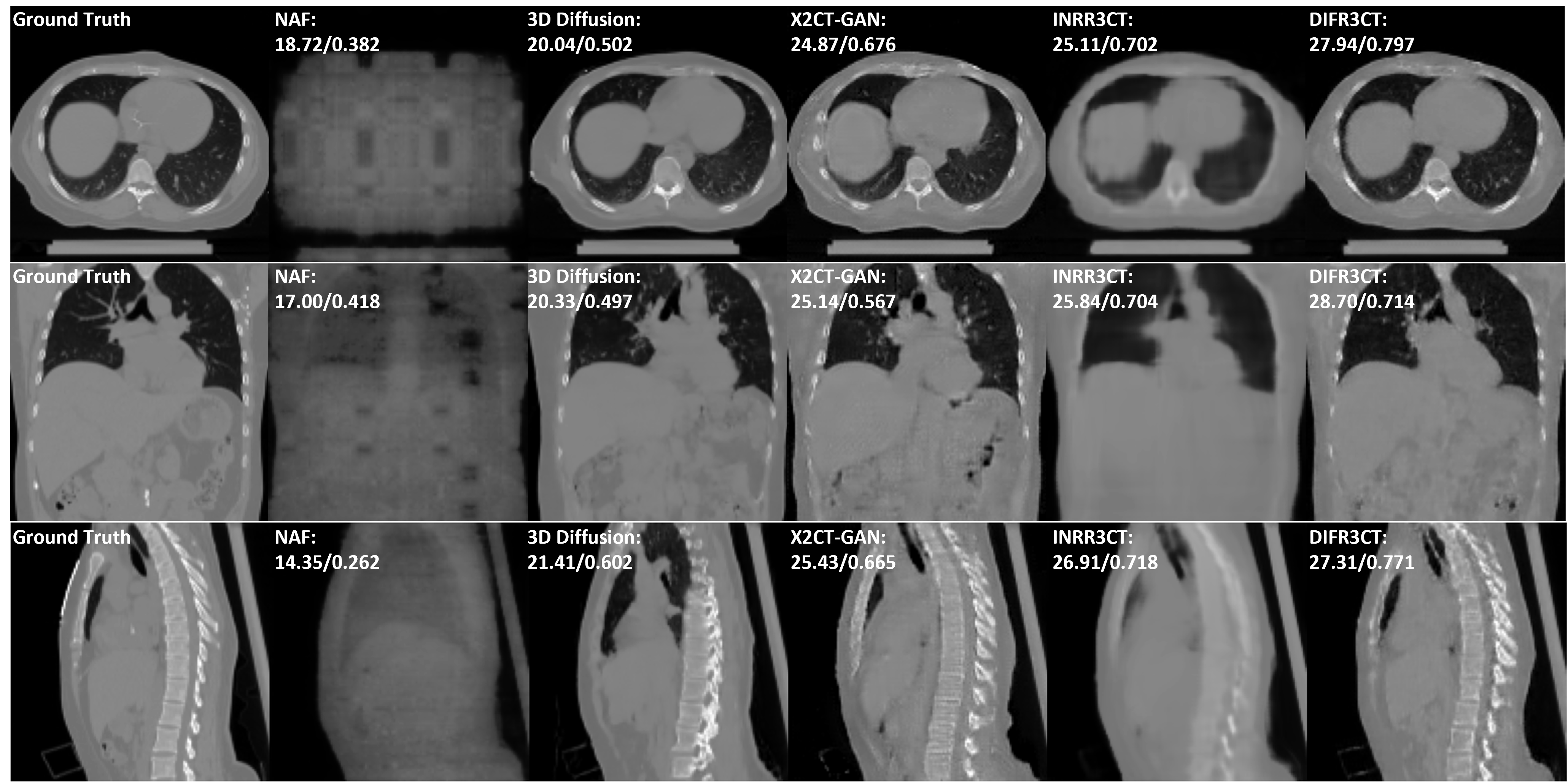}
    \caption{\textbf{Comparison of \name{} with baselines on the Thoracic Dataset, given biplanar x-ray inputs.} Each row corresponds to a different center planar view of the CT volume (axial, coronal, sagittal). The second to fifth column correspond to four baselines (marked in text on each image), and the final column shows the reconstructed 3D CT images by using the proposed \name{} method. We also report PSNR/SSIM values on each slice. DIFR3CT generates the most realistic reconstructed details of all methods.}
    \label{fig:combreast}
\end{figure*}

\begin{figure*}[t!]
    \centering
    \includegraphics[width=\textwidth]{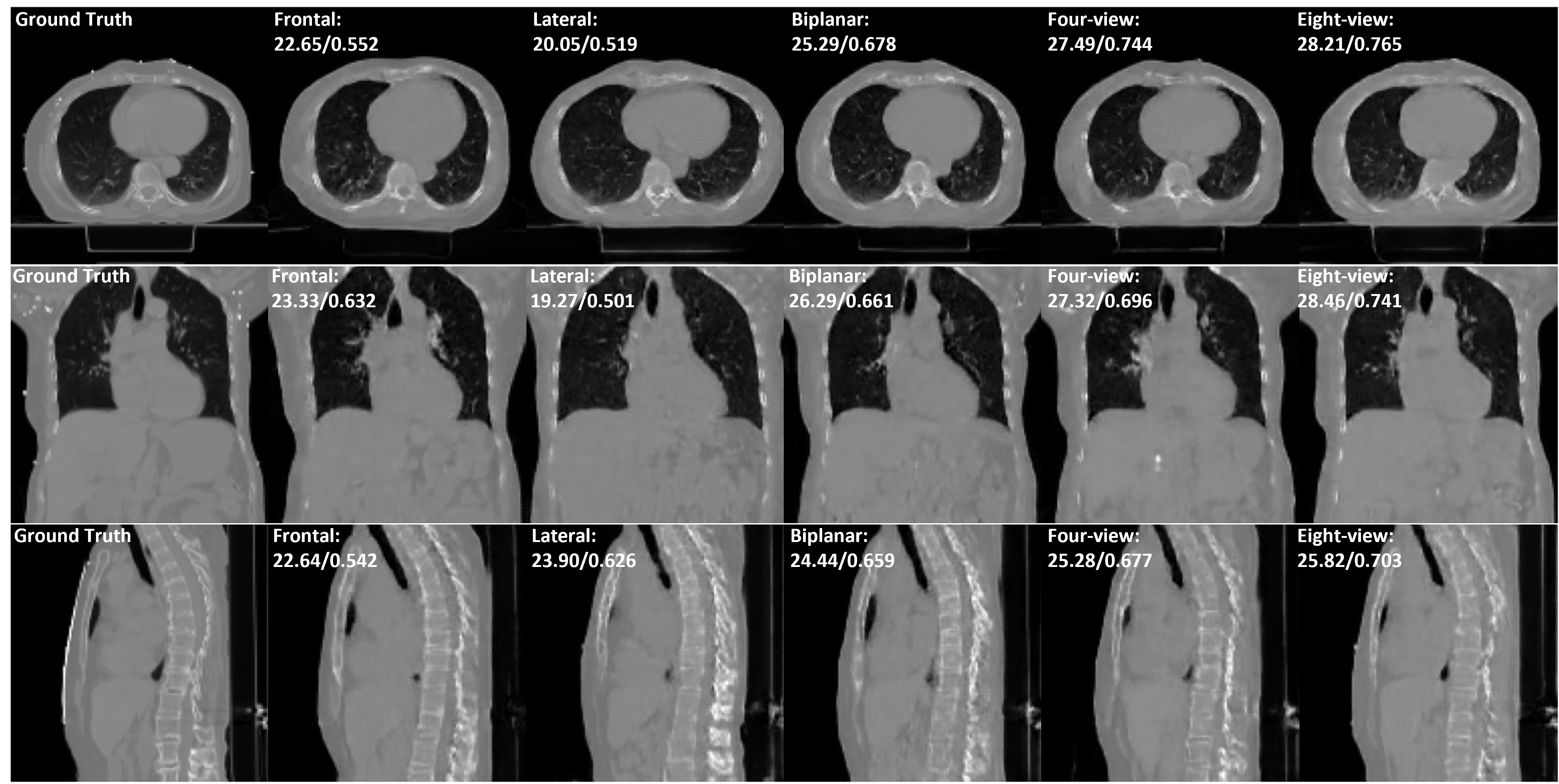}
    \caption{\textbf{Example 3D Thoracic CT reconstruction results on one patient, with varying numbers of input views.} Each row corresponds to a different center planar view of the CT volume (axial, coronal, sagittal), and the second to sixth columns correspond to a different number of views (marked in text on each image). We also report PSNR/SSIM values on each slice. As the number of input viewing angles increases, the reconstruction details improve, especially near anatomical boundaries.}
    \label{fig:breast}
\end{figure*}

\subsection{Metrics}
\label{sec:metrics}
We evaluated model performance using voxel-level reconstruction metrics (PSNR, SSIM)~\cite{wang2004image} and radiotherapy planning dose metrics. We compute PSNR and SSIM for LIDC CTs using 12-bit precision and a HU value range of $[0, 4095]$, and a HU value range of $[-1024, 3071]$ for Thoracic CTs. We compute PSNR by: $\text{PSNR}(I, \hat{I}) = 10\log_{10}(I_{\max}^2/{\text{MSE}(I, \hat{I})})$, where $I_{\max}$ is the max possible voxel value in a dataset and $\text{MSE}(\cdot, \cdot)$ computes mean squared error. We compute SSIM over local 3D windows of size $11^3$. SSIM has a range of $[0,1]$, with 1 indicating a perfect similarity.

We use common Dose Volume Histogram (DVH) metrics to evaluate radiotherapy plan accuracy: V90\% (volume of the breast receiving 90\% of the prescription dose) and V20Gy (volume of the lung receiving 20Gy). The error is defined as the difference between the DVH metrics between the plans made on the ground truth vs reconstructed CTs.  

\subsection{Reconstruction Results}
\label{sec:recon_results}
We summarize reconstruction results using PSNR and SSIM metrics for all methods on both datasets in Table~\ref{tbl:lidc-results} and Table~\ref{tbl:thoracic-results}. For 4 and 8 views, \name{} outperforms NAF and INRR3CT, the only other models capable of handling those views. For 1-2 views, \name{} outperforms all models except for INRR3CT. Although INRR3CT achieves comparable PSNR and SSIM values to \name{}, the reconstructed CTs from INRR3CT are significantly less realistic, as shown in Fig.~\ref{fig:comlung} and Fig.~\ref{fig:combreast}.

We further demonstrate visual reconstruction results by \name{} on both datasets with different number of views in Fig.~\ref{fig:lung} and Fig.~\ref{fig:breast}. As the number of input viewing angles increases, the reconstruction details improve, especially near anatomical boundaries. We observe the most significant improvement ($\sim$ 7 dB PSNR) using \name{} when moving from single to biplanar views. For the biplanar case, we provide side-by-side comparisons of \name{} with baseline methods in Fig.~\ref{fig:comlung} and Fig.~\ref{fig:combreast}. \name{} generates the most realistic reconstructed details of all methods. 3D Diffusion and X2CT-GAN introduce many artifacts, and NAF struggles to recover basic structures in the reconstructed CT images with so few inputs. INRR3CT generates blurry reconstructions lacking realism for various details (e.g., lung parenchyma, mediastinum and heart borders, and diaphragmatic surface).

\subsection{Uncertainty Quantification}
\begin{figure}[t!]
    \centering
    \includegraphics[width=\linewidth]{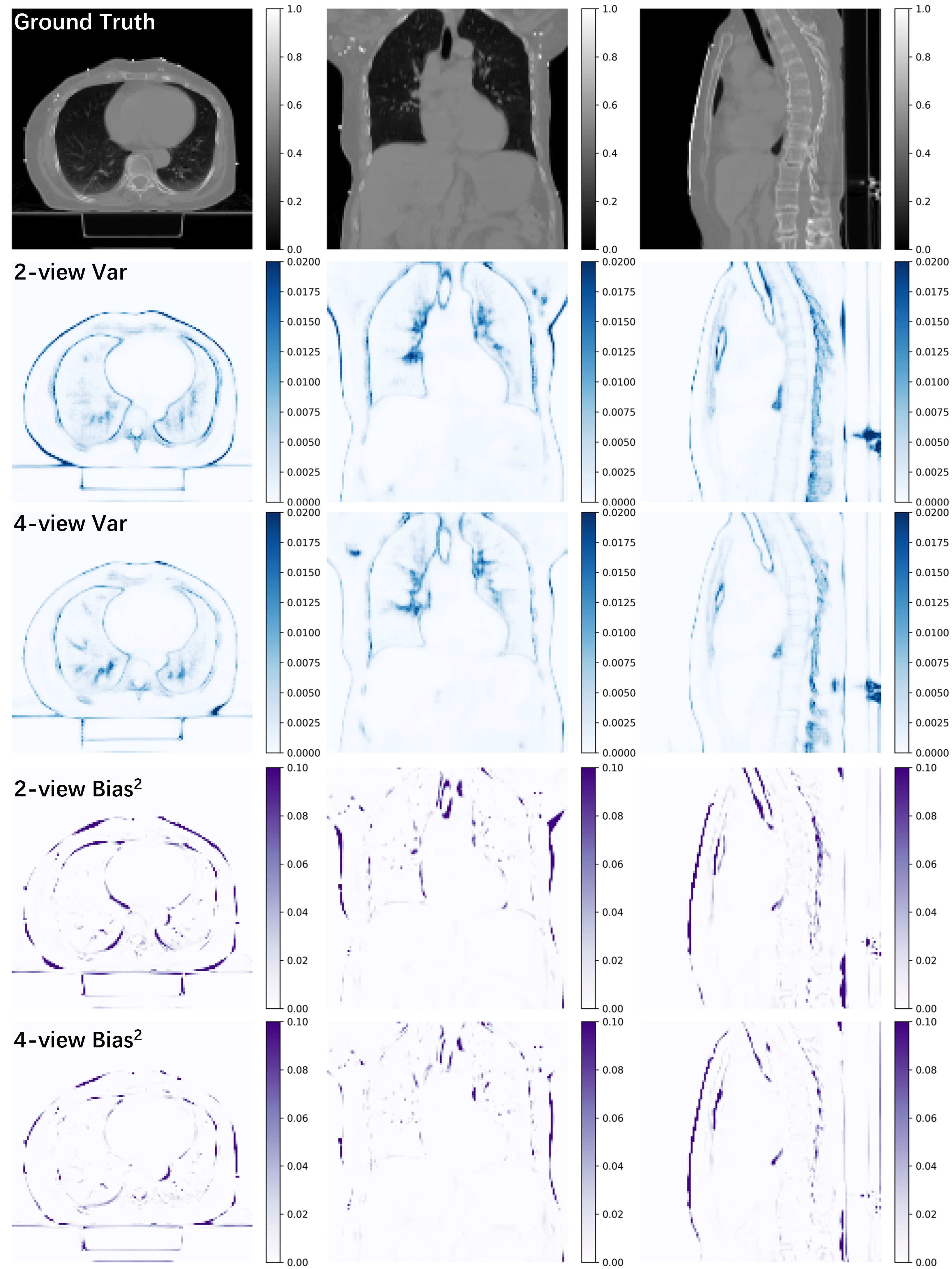}
    \caption{\textbf{Bias and variance (uncertainty) quantification visualization for one test patient from the Thoracic dataset (see Sec.~\ref{sec:uq}).} We show three center slices (axial, coronal, sagittal) of one CT scan from the Thoracic dataset (top row). Rows 2 and 3 correspond to 2-view and 4-view pixel-wise variances using reconstructions from \name{}, while Rows 4 and 5 correspond to 2-view and 4-view pixel-wise squared bias using reconstructions from \name{}. Darker blue and purple values indicate higher values of variance and bias.} 
    % \GB{Give takeaway message -- do the uncertainties make sense?}
    \label{fig:uncertain}
\end{figure}

We next demonstrate per-voxel variance (aleatoric uncertainty) and bias estimation for a single patient, over multiple predicted CT samples (see Sec. \ref{sec:uq}). We evaluated two pretrained \name{} models using 2-view and 4-view inputs. Fig.~\ref{fig:uncertain} presents variance and squared bias maps for one patient over 100 MC samples. Darker shades of blue/purple indicate higher variance (uncertainty) and bias. Contours of organs and bones have particularly high bias and uncertainty, which is reasonable since these regions have large spatial image gradients. Strong contours between air (black) and bone (white) are particularly prone to high bias values, if \name{} systematically makes geometrical errors in those regions. There are particularly high variance values around the lung bronchioles, which is reasonable since they are thin structures with less predictable patterns. Additionally, \name{}'s uncertainty for these areas significantly reduces as the number of input views increases from 2 to 4 views. 

\subsection{Radiotherapy Treatment Plan Case Study}
\label{sec:radiotherapy_results}
Finally, we performed a case study exploring \name{}'s viability for radiotherapy (RT) planning. We first generated automatic whole breast RT plans from ground truth CTs of 5 test patients with intact breast tissues. We use these patients for testing instead of those in Thoracic, because automated breast RT models generally assume intact breast tissues. We generated RT plans using synthetic CTs generated by \name{} given 8 input x-ray views, and computed dose volume histogram metrics on contours from both the synthetic and ground truth RT plans. 

\begin{figure}[t!]
    \centering
    \includegraphics[width=\linewidth]{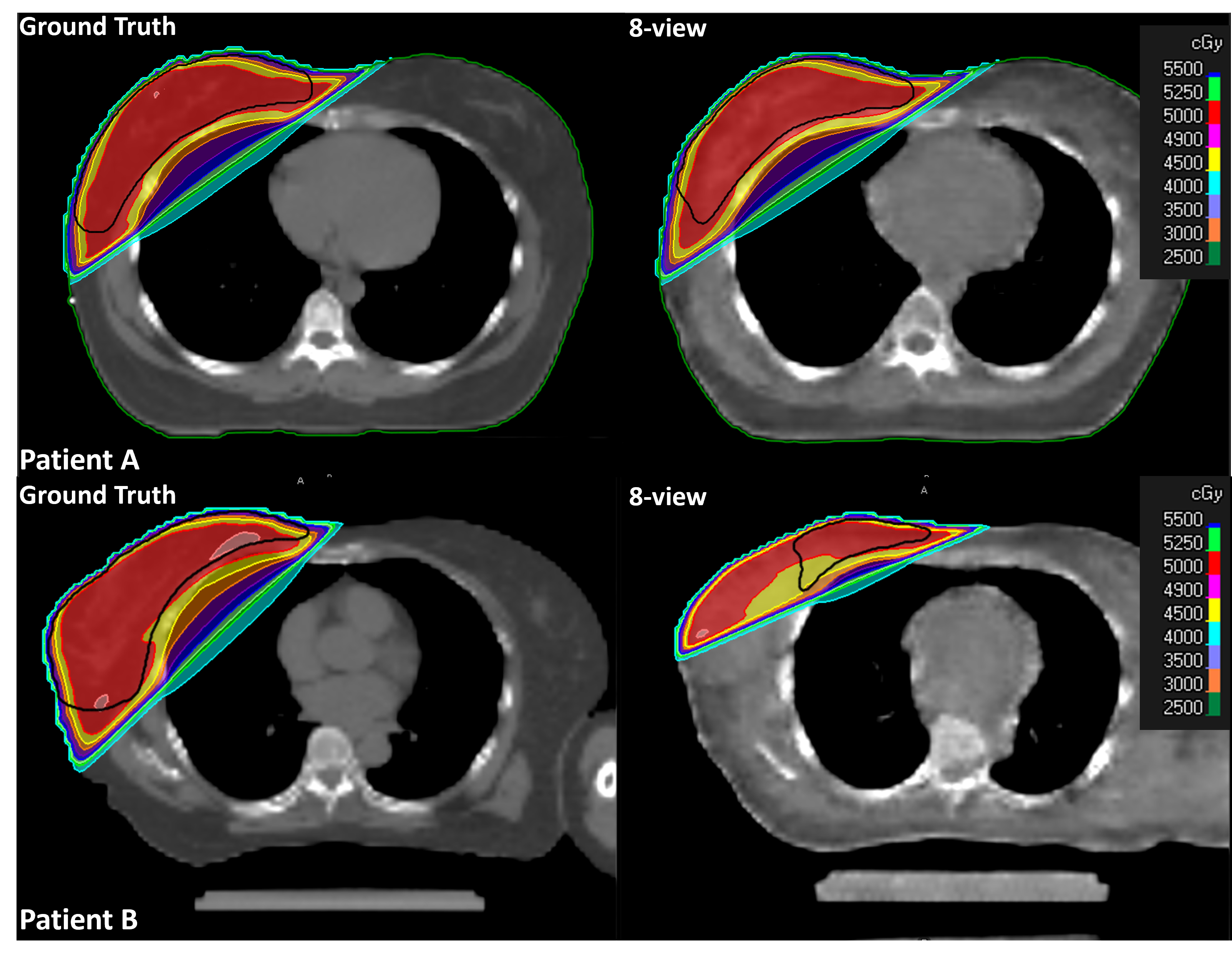}
    \caption{\textbf{Dose distributions for two-field opposed beam radiotherapy treatment plans for the breast, using ground truth CTs (left) and 8-view reconstructed CTs from \name{} (right).} Shown are axial reformations of the given CTs (gray) with overlaid dose distributions (color). The automated contour of the breast is traced with a black line. The CT synthesized by \name{} for Patient A (top) yields an acceptably close dose distribution to ground truth, while the one synthesized for Patient B (bottom) does not.} 
    \label{fig:rt}
\end{figure}

Reconstructions for three out of five patients produced sufficient whole breast contours, permitting RT planning. For these three plans, V90\% and V20Gy were clinically acceptable. The error in V90\% ranged from 1-9\%, and the error in V20Gy ranged from 0-46\%. Fig.~\ref{fig:rt}-top presents an RT plan visualization for one of these three patients. Reconstructed CTs on the remaining two patients did not have sufficient soft tissue contrast to automatically segment the breast, and automated planning was not clinically acceptable. For these patients, error in V90\% ranged from 18-43\%, and the error in V20Gy ranged from 86-99\%. Fig.~\ref{fig:rt}-bottom presents an RT plan visualization for one of these two patients.

\section{Discussion and Conclusion}

In this work, we propose \name{}, the first conditional latent diffusion model for high-quality 3D CT scan reconstruction from extremely few planar x-ray views. \name{} enables the recovery of high-quality CT images that preserve geometric structure and sharp edges by utilizing a latent diffusion model conditioned on fused features from given different-angle planar x-ray images.

\name{} outperforms various baselines in terms of PSNR/SSIM and visual inspection on both the LIDC and Thoracic datasets. \name{} is also the only model that is both flexible and computationally scalable with increasing numbers of x-ray views. For example, while INRR3CT~\cite{sun2023ct} and 3D Diffusion~\cite{ho2022video} can work with an arbitrary number of views, they required significantly more memory during training than \name{}, limiting their practicality for 4 or more input views. Additionally, diverse realizations produced by \name{} enable meaningful uncertainty quantification via Monte Carlo sampling. Such sampling is not possible for deterministic baselines such as X2CT-GAN~\cite{ying2019x2ct} and INRR3CT~\cite{sun2023ct}.

Results from the RT case study are promising, demonstrating that for three out of five patients, \name{} can produce RT plans nearly identical to those produced from ground truth CT. This suggests that \name{} has potential for use in clinical settings where CTs are unavailable. Due to data availability, we trained \name{} on patients who had mastectomy, and tested on patients with intact breast tissue. This distribution mismatch will certainly introduce error. A more in-depth clinical study is needed to understand how this and other factors impact reconstruction reliability.

As in virtually all prior works, one clear limitation of our study is that we developed and evaluated our models primarily on synthetic x-rays produced from DRR generators. This strategy ensures that CT and x-ray pairs are perfectly aligned, and that x-rays are precisely acquired. However, real x-rays have different resolution and noise properties from DRR generations, and these properties can slightly vary from one machine to the next. In addition, in clinical practice, planar x-rays will not be acquired at perfectly precise orientations. Important next steps include developing and evaluating reconstruction models that may be trained on a combination of synthetic and real x-rays, and that can handle uncertain acquisition factors at test time.

Finally, while the inference power of deep neural networks is remarkable, they are also known to \emph{hallucinate} details that look realistic, but are incorrect. Further analysis is needed to understand how reconstruction models such as \name{} hallucinate, particularly on atypical patient cases underrepresented in the training data. Uncertainty estimates can help flag likely hallucinated reconstructions, but only if the uncertainty bounds are properly \emph{calibrated} with ground truth data~\cite{vovk2020conformal}, and analyzed with respect to out-of-distribution test cases. 

%Finally, most deep neural network reconstruction studies, including ours, assume image data have fixed sizes. In this study, we resamp x-ray and CT pair datasets as squares and cubes for training and testing via resampling or zero-padding. These operations can change the original patient body structure or result in memory overhead. Extending the flexibility of diffusion reconstruction models to handle data with arbitrary dimensions would be a valuable contribution.

\appendices
% \section*{Appendix}
% \section*{Acknowledgment}
% \textbf{Author contributions:} The project was conceived by Y.S. and G.B.. The project was primarily supervised by G.B., assisted by A.V.. The clinical parts were primarily supervised by T.N. and L.C., assisted by O.M.. The methods were developed by Y.S.. The code of the model was implemented by Y.S.. The numerical results were collected by Y.S.. The automated breast radiotherapy contouring and planning was implemented by H.B.. The data acquisition and preparation was conducted by T.N., L.C., Y.S. and H.B.. The manuscript was primarily drafted by Y.S., assisted by H.B. and T.N.. The manuscript was edited by G.B.. All authors have reviewed the manuscript. 

% \textbf{Competing interests:} The authors declare no competing interests.

% \textbf{Data and materials availability:} The data and code used for reproducing the results in the manuscript is available at \href{https://github.com/yransun/DIFR3CT}{https://github.com/yransun/DIFR3CT}.

\bibliographystyle{IEEEtran}
\bibliography{refs} 

\end{document}